\documentclass[journal]{IEEEtran}
\usepackage{ifpdf}
\usepackage{cite}
\usepackage{graphicx}
\usepackage{color}
\graphicspath{{fig/}}
\usepackage{amsmath}
\usepackage{array}
\begin{document}

\title{Long Working Distance Apodized Grating Coupler}
\author{Md~Shofiqul~Islam~Khan, Sylvain~Combri\'{e} and~Alfredo~De~Rossi,~\IEEEmembership{Member,~IEEE}
\thanks{M.S.I. Khan, S. Combri\'{e} and A. De Rossi are with Thales Research \& Technology, Palaiseau, France.}
\thanks{Manuscript received XXXX}}

\markboth{arxiv,~2020}%
{Shell \MakeLowercase{\textit{et al.}}: Bare Demo of IEEEtran.cls for IEEE Journals}
\maketitle

\begin{abstract}
We design a focusing grating coupler by a simultaneous apodization of the filling factor and the period. In addition to in plane focusing to the input waveguide providing a total length of less than 70 $\mu m$, a further apodization of the curvature allows out of plane focusing into a fiber set $150 \mu$m away from the grating surface. The design is proposed for novel semiconductor on insulator waveguides such as GaInP. The coupling efficiency is calculated from two dimensional simulation which is about $45\%$.
\end{abstract}


\IEEEpeerreviewmaketitle

\section{Introduction}
\IEEEPARstart{C}{oupling} between a fiber and an integrated waveguide is a challenging task due to their vastly different sizes, standard optical fiber (SMF-28) is about two orders of magnitude larger compared to a typical wire waveguides. This mismatch can amount to less than $- 20$ dB coupling efficiency (for nanowires of 500 nm x 220 nm cross section), when the two waveguides are butt-coupled for some applications which is unacceptable. A surface grating, coupling light propagating out of plane (vertically or under a small tilt angle) from a standard fiber to any integrated waveguide on the chip was demonstrated in 1970 \cite{dakss1970grating}. The emergence of Silicon photonics, namely the silicon-on-insulator (SOI) platform compatible with established integration and packaging technologies, has motivated the development of very efficient grating coupler\cite{taillaert2002out}. This allows a wafer scale testing with simpler fiber adjustments and good alignment tolerances for very low loss penalty in lateral and longitudinal directions \cite{taillaert2003efficient}.\\
It has been demonstrated recently that diffraction grating coupler can reach a theoretical efficiency as high as $-0.5$ dB, still maintaining a large bandwidth of 35 nm \cite{taillaert2004compact}, which has been possible \cite{Andreani2015} using a thicker Si layer than the conventional 220 nm thick SOI waveguide provided by silicon photonics foundry. However, this design is based on a sophisticated two dimensional pattering requiring a complex design and accurate fabrication, which is certainly justified when the performance is to be maximized. In this manuscript we rather consider a simpler design and we restrict on a geometry consisting in curved etched lines which can be described by a limited set of parameters.\\ 
The coupling efficiency of the simplest grating structure, consisting of a strictly periodic structure, faces a theoretical limit around $-7$ dB \cite{bogaerts2004basic}. The performance has been quickly enhanced by introducing some changes such as: dual grating-assisted directional coupling \cite{masanovic2005high}, gold mirrors \cite{van2007compact}, distributed Bragg reflectors (DBR) \cite{selvaraja2009highly}, back reflectors \cite{zaoui2012cost}, high index contrast grating \cite{bekele2015polarization} and apodized gratings \cite{chen2010apodized,mekis2011grating,mekis2012cmos,can2013high}. In all these cases, these structures have a very short working distance ($< 5 \mu$m). In some cases the fiber cannot be brought close enough \cite{oton2016long}, therefore lens-like (long working distance) grating coupler comes into play.\\
	\begin{figure}[ht!]
    	\centering
 		\includegraphics[width= 2.5 in]{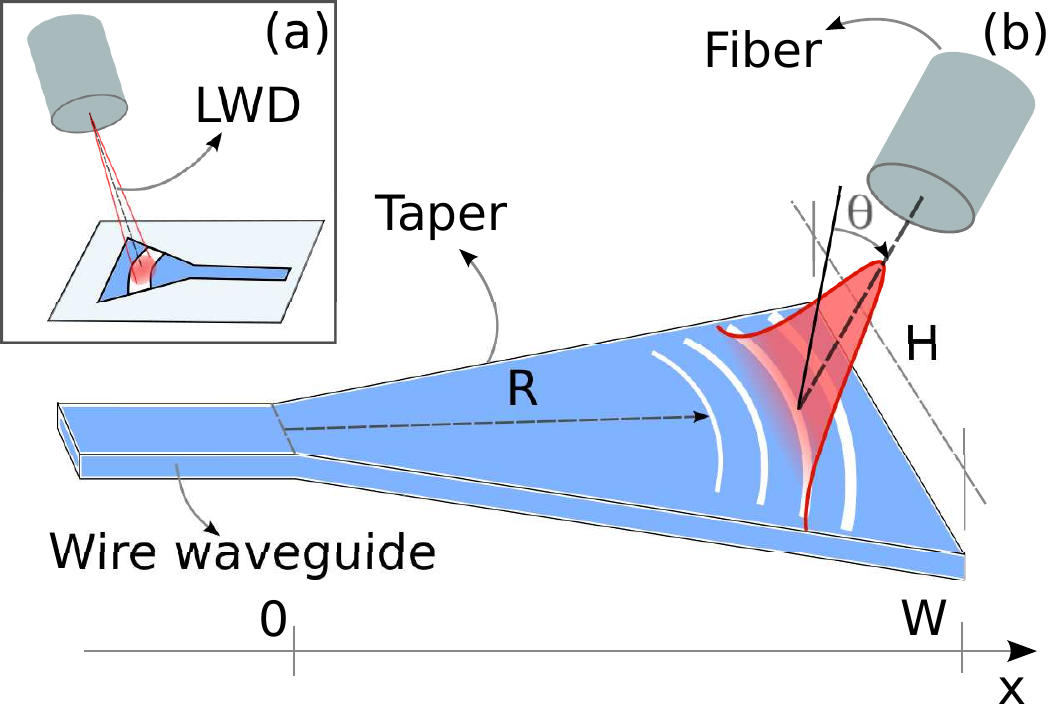}
		\caption{(a) Long working distance (LWD) grating coupler (inset). (b) In-plane focusing taper for diffracting circular wavefront.}
		\label{fig:mode_match}
	\end{figure}
Lens like grating coupler (fig. \ref{fig:mode_match}) focuses the out coupled light on the fiber located at a distant place without the presence of any external lens and can be implemented by a suitable modification of the grating stripes. More precisely, the phase front of the wave emerging from the grating is shaped to create a converging Gaussian beam focused into the fiber input\cite{oton2016long}. Such design was implemented on uniform grating stripes which were bent such a way that the diffracted wavefront matches the phase of a wave coming from a distant fiber. Attempt of incorporating apodization would add computational complexity into this design. Whereas, we show that focusing and very good coupling efficiency can be achieved with a design-wise simpler approach where only the grating periods are modulated with slight deformation in the grating stripes. In our design, focusing effect is independently achieved in the direction of propagation and in its transverse direction using simple transformation equations which are equally applicable for uniform and apodized gratings.\\
In this article, section \ref{sec:apdization}, illustrates design method of apodized grating, introduce to device structure and modeling.  Based on that apodized grating design section \ref{sec:focusx} and \ref{sec:focusy} would further proceed illustrating the design procedure for beam focusing and eventually summarizing modeling outcomes for lens like grating coupler in section \ref{sec:focusboth}.

\section{Apodized Grating Coupler Design}
\label{sec:apdization}

\begin{figure}[ht!]
		\centering\includegraphics[width= 3 in]{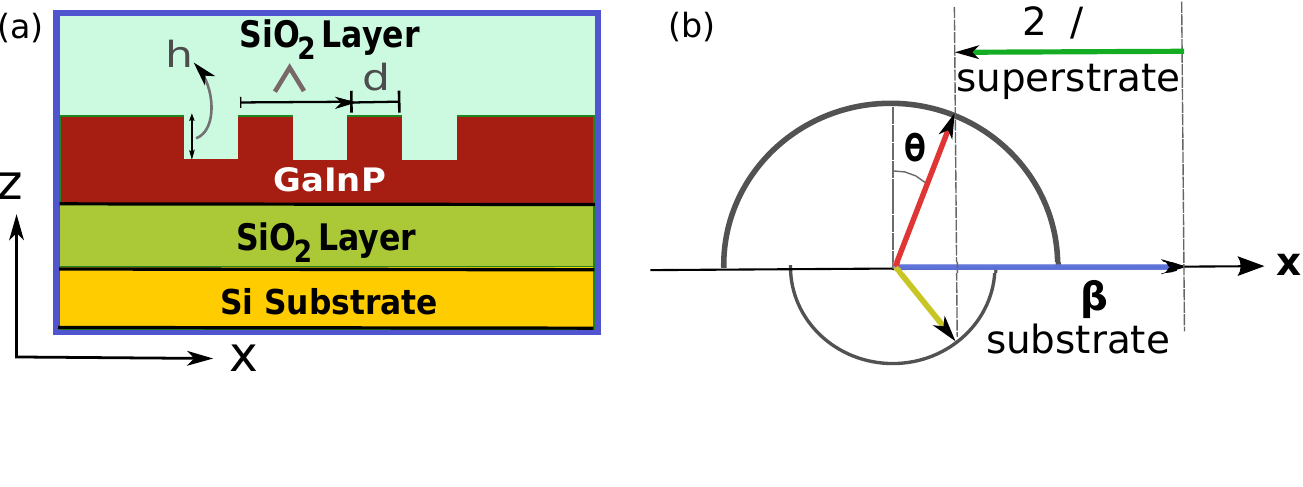}
		\caption{(a) Example of the Cross section ($x-z$ plane) of the structure used in the CAMFR calculations.  (b) first order Bragg scattering .}
		\label{fig:bragg_cond}
		\end{figure}

Apodization can be implemented in many ways, here we restrict the choice to two parameters, the period $\Lambda$ and the filling factor $\phi=d/\Lambda$  (Fig.\ref{fig:bragg_cond}).  These parameters are easily controllable by design, in contrast with the depth of the etch\cite{bates1993gaussian} , which is fixed. For the purposes of one example, we consider here Gallium Indium Phosphide on silicon, which has been developed recently for on-chip nonlinear photonics\cite{dave2015nonlinear,martin2017}.
The grating operates in the  first diffraction order  (figure \ref{fig:bragg_cond}(b)), hence:
		\begin{equation}
		\beta - K_g = K_x
		\label{eqn:bragg_condition}
		\end{equation}
	where $\beta = 2\frac{\pi}{\lambda}n_{eff}$ is the propagation constant of the guided mode, $K_g = \frac{2\pi}{\Lambda}$ and the projected wave vector of the incident mode is $K_x = \frac{2\pi}{\lambda}\sin\theta$, with $\theta$ as the angle of incidence (fiber axis) with respect to the surface normal and $\lambda$ as the wavelength of operation. 
In a nonuniform grating, the leakage factor $\alpha$ is a function of position $x$ which can be manipulated to obtain a desired output beam. $\alpha (x)$ is related \cite{waldhausl1997efficient,bates1993gaussian}  to the output beam shape as
	
	\begin{equation}
			\alpha (x) = \frac{G^2(x)}{2 [1 - \int ^x_0 G^2(t) dt]}
			\label{eqn:leakage_factor}
	\end{equation} 
where $G(x)$ is the target beam profile, hence  a normalized Gaussian profile. The target $\alpha(x)$ is achieved by mapping it into values of the filling factor, as in \cite{waldhausl1997efficient}. Very importantly,  the change of the filling factor involves a change of the effective index, which has to be compensated by a change of the period in order to keep the wavevector of the emerging field constant. 

\begin{figure}[ht!]
	 		\centering\includegraphics[width=0.8\linewidth]{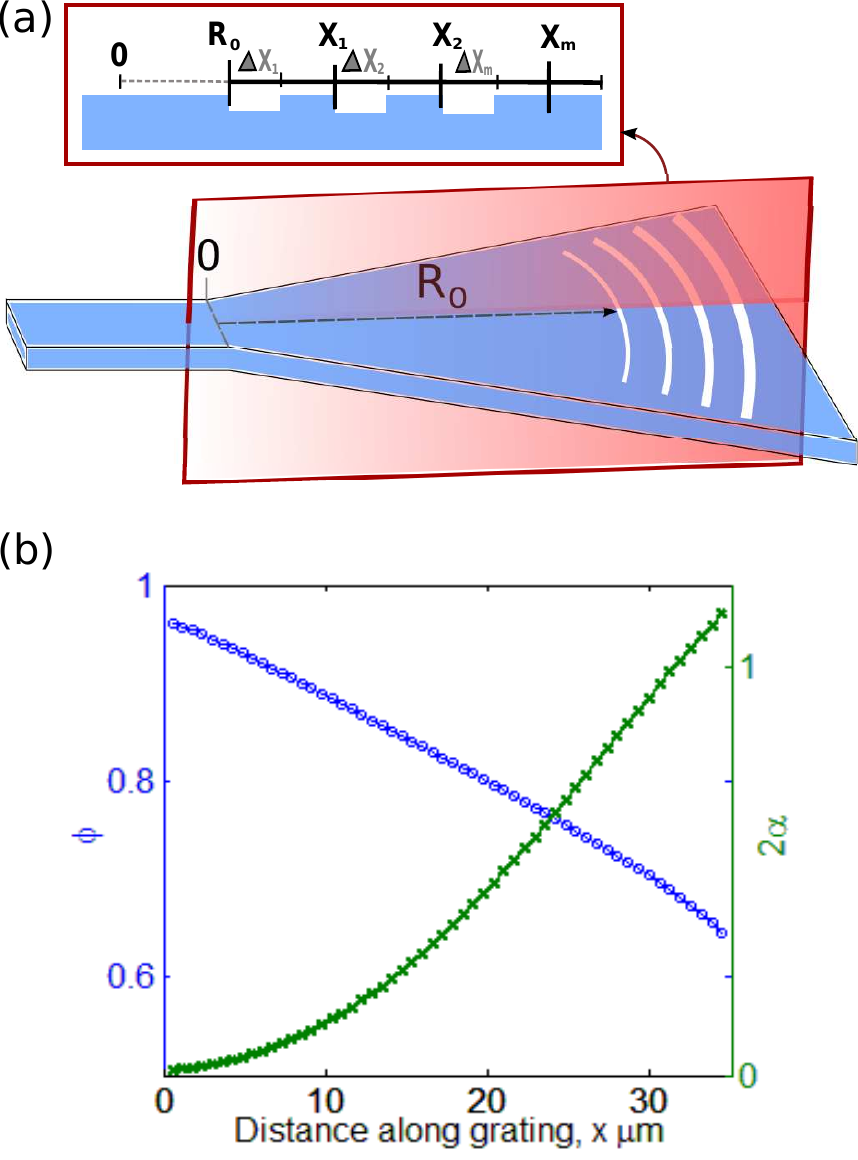}
	 		\caption{(a) In-plane focusing grating coupler and (b) apodized grating parameters.}
	\label{fig:grating_parameters}
\end{figure}	

We first design a linear apodized grating. Hence, we use the Cavity Modeling Framework (CAMFR) \cite{bienstman2008camfr}. An example code from \cite{taillaert2004grating} was adapted and modified for the modeling of our grating designs. More precisely, fixing the wavelength, the etching depth and the scattering angle ($\theta$) we compute the leakage factor as a function of the filling factor. In this process, an iterative procedure is used to adjust the period $\Lambda$ accordingly. As a result, a look-up map $[\alpha,\Lambda] =F(\lambda,\phi,\theta)$ is built. This is used to map the spatial dependence of $\alpha$ into the profile of $\phi$ and $\Lambda$, which we use to build the linear grating (cross sectional details in fig. \ref{fig:grating_parameters}(a)) using the  straightforward rule for the position of the m-th etched line: with width $\Delta x_m=\Lambda_m(1-\phi_m)$ and edge $x_m =\sum_{l=0}^m \Lambda_l$. We have chosen an  etch depth of 140 nm, namely roughly half of the thickness (290 nm) of the GaInP guiding layer with a oxide layer of 1 $\mu$m. Obtained apodized grating parameters are presented in fig. \ref{fig:grating_parameters}(b). \\
The next step consists in converting the linear grating into an in-plane  focusing grating. This is achieved by mapping the linear grating described by the position and width of the etched lines {$x_m$, $\Delta x_m$} into arcs with center set at the waveguide end (fig. \ref{fig:grating_parameters}(a)) and radius $R_m=x_m + R_0$ (where $R_0 \gg$ Rayleigh length).

\subsection{Grating simulation through 3D calculations}

\begin{figure}[ht!]
	 		\centering\includegraphics[width=\linewidth]{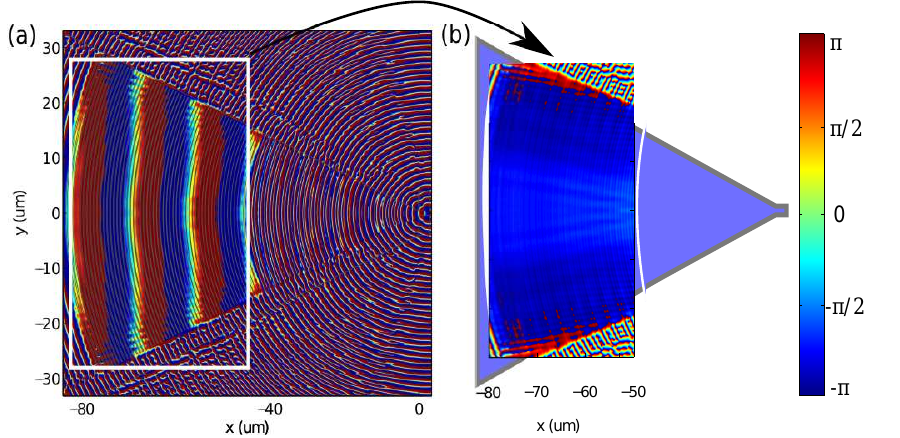}
	 		\caption{(a) Phase $\varphi(x,y)$ of the electric field  ($E_y$) over the grating and (b) detail of the area over the grating coupler showing $\varphi-2\pi x \sin(\theta)/\lambda$.}
	\label{fig:FDTD_focusing_apodized}
\end{figure}

In-plane focusing taper is designed (as in fig. \ref{fig:grating_parameters}) that accommodates circular grating stripes and converts the diffracted beam size to a circular beam shape. Maximum length of the taper is W $= 80 \mu$m and width is H$= 64 \mu$m. The focusing grating is simulated using the 3D FDTD method implemented with an in-house code running on a parallel computer server (Xeon E2610, 12 cores, 24 threads). This code has been extensively used for the design of photonic crystals devices\cite{NVQTran2011}.\\
In the designed structure of in-plane focusing grating coupler, only GaInP guiding slab sandwiched between oxide layers is taken into account for 3D FDTD simulation in order to ensure reduced simulation time. The purpose of such 3D simulation was to extract output beam shape, intensity distribution, phase of the diffracted light etc. At $\lambda = 1550$nm, phase profile of the magnetic field component $H_x$  is presented in fig. \ref{fig:FDTD_focusing_apodized}(a). Subtracting the linear terms from this phase profile should provide a flat phase front (constant value) above the grating which is evident in fig. \ref{fig:FDTD_focusing_apodized}(b), nearly flat phase front is observed with a hump in the phase profile along the y-direction. This phase modulation is introduced by the bending of the grating stripes. Further investigation shows that perfect flat phase front can be achieved by shifting the origin of the grating stripes by several microns (not presented here).

\begin{figure}[ht!]
 	 \centering
 	 \includegraphics[width=\linewidth]{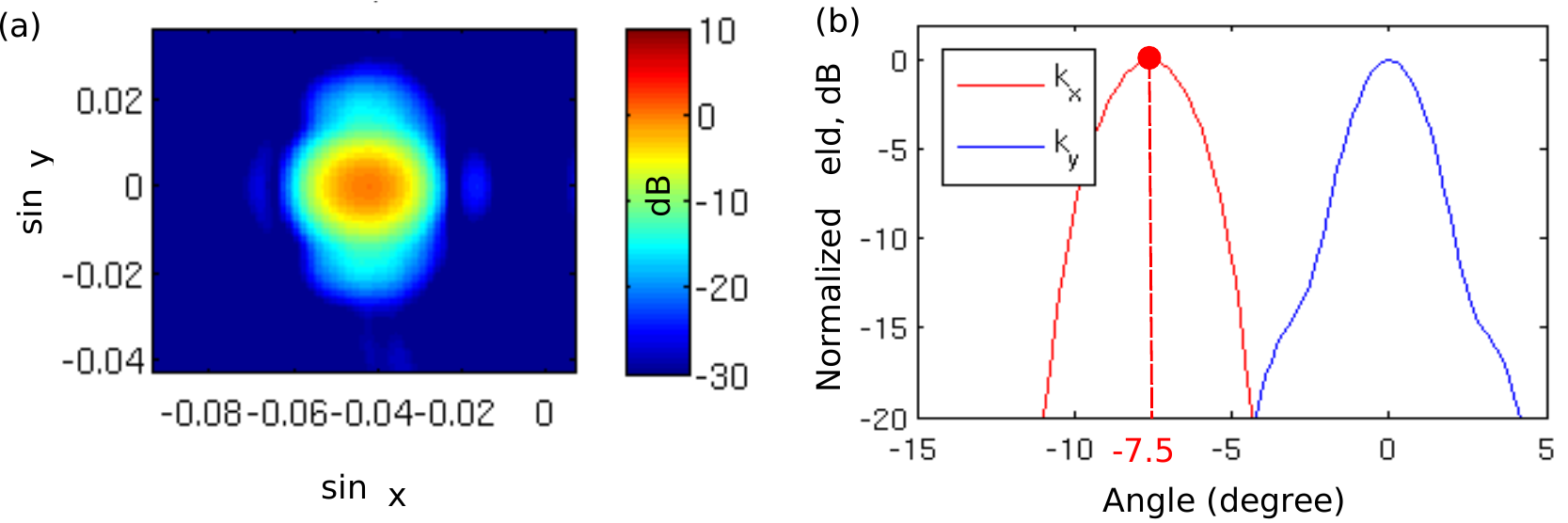}
	 	\caption{Normalized real value of Poynting vector for diffracted light in the upward direction (Fraunhofer 
approximation) at $\lambda = 1550 nm$ measured $0.7 \mu$m above the grating surface. (a) Two dimensional false color map and (b) cross cut over $k_y=0$ and $k_x=2\pi/\lambda 0.045$.}
	\label{fig:Lens_farfield_apo_b}
\end{figure}

Fig. (\ref{fig:Lens_farfield_apo_b}) shows the square of the spatial Fourier transform of the field components parallel to the grating surface (at $z = 0.7 \mu$m), which is proportional to the power flux in the far field (Fraunhofer approximation). Spatial Fourier transform of the diffracted beam measured is presented in fig. (\ref{fig:Lens_farfield_apo_b}). Beam width extracted along x-axis is 0.050 radians and along y-axis is 0.040 radians. Maximum efficiency is expected to be detected at an angle of 7.5 degree with respect to the axis perpendicular to the grating surface.


\section{Focusing along X Direction}
\label{sec:focusx}
The idea of focusing the output beam in the direction of propagation is inspired by the propagating Gaussian beam. Except for the near vicinity of the source traveling Gaussian beam has parabolic phase front. We intend to modify the grating periods such a way that it modulates the diffracted phase front in a parabolic fashion as if it is coming from a Gaussian beam. It is anticipated that such bending of the diffracted phase front would focus the output beam at some distance.   

	\subsection{Propagating Gaussian Beam}	 
	While propagating in a homogeneous medium Gaussian beam solution can be expressed as \cite{yariv1975quantum}:
	
	\begin{equation}
	E \propto \frac{1}{R} e^{-ikR} \simeq \frac{1}{R} exp (-ikz - ik \frac{x^2+y ^2}{2R})
	\label{eqn:gauss_prop_simplified}
	\end{equation}
	
	where $x^2 + y^2 \ll z^2$ and z is equal to R. Gaussian beam's radius of curvature $R = z ( 1 + \frac{z_0^2}{z^2} )$ where $z_0 \equiv \frac{\pi w_0^2 n}{\lambda}$. Here, $n = $ refractive index of the medium, $k =$ wave vector, $\lambda = $ wave length and $r$ is the radial distance on the x-y plane of the beam . It follows from the eqn. (\ref{eqn:gauss_prop_simplified}) that, except for the immediate vicinity of $z = 0$, the wavefronts are parabolic since they are defined by $k[z+(r^2/2R)] = constant$. We intend to obtain such parabolic phase in the diffracted beam of grating and presume a focusing effect out of plane far from the grating surface.

	 \subsection{Focusing along the Direction of Mode Propagation} 
	 \label{subsec:Lens_grating}
	 
	 For an uniform grating, phase of the diffracted field can be retrieved from the Bragg's condition (eqn. \ref{eqn:bragg_condition}) as follows:
	 
	 	\begin{equation}
			 \phi = \frac{2 \pi}{\lambda} ( n_{eff} - \sin \theta ) x= \frac{2 \pi}{\Lambda} x = K
			 \label{eqn:phase_uniform}
	 	\end{equation}
	 
	 In case of an apodized grating the diffracted field has plane phase front. As compared to the eqn. (\ref{eqn:gauss_prop_simplified}), an addition of a term to eqn. (\ref{eqn:phase_uniform}) that varies in a parabolic fashion with the distance along the grating would bend the linear phase front of the apodized grating. Considering K as a linear spatial variant, we add $\Delta x$ variation to the grating periods such a way that it introduces a parabolic modulation to the grating units. After the phase modulation the phase front can be expressed as:
	 
			 \begin{equation}
			  		\phi' = \frac{2 \pi}{\lambda} (\sin \theta) x + \Delta \phi = \frac{2 \pi}{\lambda} ( (\sin \theta) x - \frac{x^2}{2R} )
			  		\label{eqn:phase_1}
			 \end{equation}
			 
			 and it can also be written as
			 
			 \begin{equation}
					\phi' = ( \frac{2 \pi}{\lambda} n_{eff} - K ) x 
					\label{eqn:phase_2}
			 \end{equation}
			 
	where $\Delta \phi =  \frac{2 \pi}{\lambda} n_{eff} \Delta x = - \frac{2 \pi}{\lambda} \cdot \frac{x^2}{2R}$. We can intuitively define $\Delta x = - \frac{x^2}{2Rn_{eff}}$.	Hence, if the distance along the grating is x then for parabolic phase modulation it will become, $x' = x + \Delta x = x (1 - \frac{x}{2R n_{eff}}) $. 	 
			 
	\vspace{0.5em}

\begin{figure}[ht!]
	 	 	 		\centering\includegraphics[width=\linewidth]{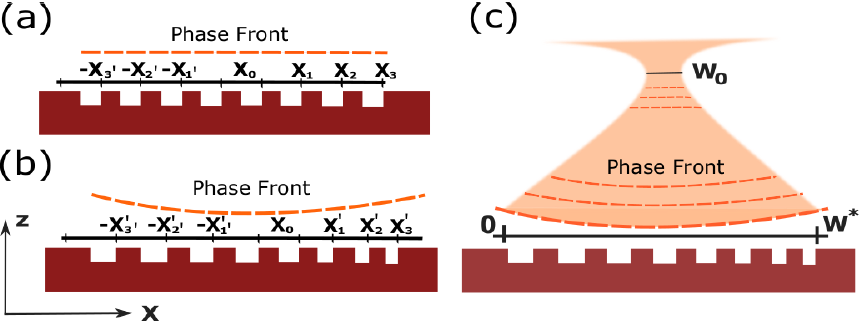}
	 	 	 		\caption{Modulating the grating periods along the direction of mode propagation for obtaining focusing phase front. (a) Grating before phase modulationm (b) stretching and compression of periods for obtaining phase modulation, and (c) focusing phase front after modulation.}
	 	 	 		\label{fig:focusx}
	 	 	 	\end{figure}

	 To understand the concept of phase modulation of the grating coupler, we start with a grating structure which is diffracting light vertically (figure \ref{fig:focusx} (a)). $X_0$ is the period associated with the beam center of the diffracted Gaussian field. $X_{3'}, X_{2'}, X_{1'} $ and $ X_1, X_2, X_3$ are the distances from the central period. Focusing phase front can be achieved by the following transformation of the existing grating structure
	 
	 \begin{equation}
	 x' = x (1 - \frac{x}{2R_x n_{eff}})
	 \label{eqn:focusx}
	 \end{equation}   
	 
	 where $R_x$ is the radius of curvature for the parabolic phase front and $n_{eff}$ is the effective index of the propagating mode. After such transformation periods on the left of the central period are stretched ($X'_{n'}$ \textgreater $X_{n'} $) and periods to the right are contracted ($X'_{n}$ \textless $X_{n}$) as depicted in figure \ref{fig:focusx} (b).     
	  
Such transformation adds up a parabolic term in the phase of diffracted beam by the grating and by modulating the phase with proper radius of curvature that beam can be focused within few hundred microns above the grating (fig. \ref{fig:focusx}(c)) surface for long distance operation. 	 	 	 	
	 	 	 	
\subsection{Simulation Results: Focusing along X}
\label{subsec:sim_res_focusx}
	
\begin{figure}[ht!]
 \centering
	 \includegraphics[width=2.0in]{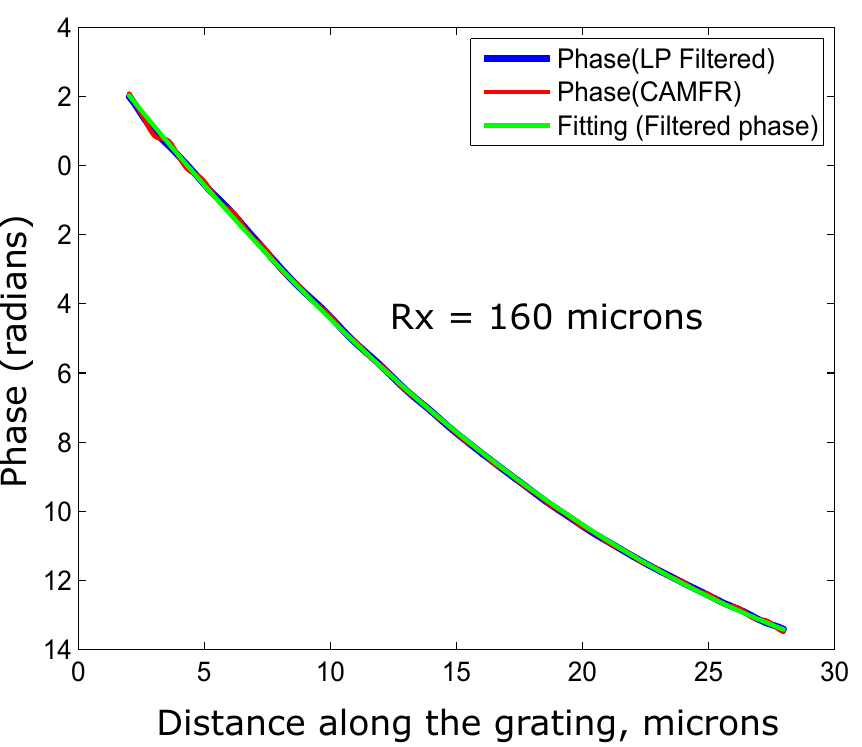}
 		\caption{Phase profile of the $H_x$ component measured at $z = 0.7 \mu$m above the grating coupler and radius of curvature calculated $R_x = 160 \mu$m. Calculated phase by CAMFR was low-pass filtered to remove evanescent modes from the phase profile.}
  \label{fig:Lens_focusx_camfr}
\end{figure}

As discussed in the section \ref{subsec:Lens_grating}, phase modulation with a radius of curvature $R_x = 150 \mu$m is applied to the apodized grating parameters ($\Lambda$, FF) in the x-direction. In fig. (\ref{fig:Lens_focusx_camfr}), 2D CAMFR simulation shows second order polynomial fit to the phase profile of the diffracted beam, the radius of curvature is extracted after spectral filtering diffracted phase which tends to be $\sim 160 \mu$m close to our design value. 

It was observed that radius of curvature of the phase front follows the design (input) value more closely if varied between 200 and 500 $\mu$m. The phase profile shows the phase extracted from CAMFR simulation (red) which is then filtered to get rid of evanescent modes which makes fitting difficult resulting in erroneous extraction of radius of curvature. Low pass filtered phase profile (blue) is then fitted to a second order polynomial (green) from which the radius is determined. If second order coefficient is $P_2$ then from eqn. (\ref{eqn:phase_1}) radius of curvature can be deduced as:

	\begin{equation}
		 R = \frac{\pi}{\lambda P_2}
		 \label{eqn:radius_curvature}
	\end{equation}

\section{Focusing along Y Direction}
\label{sec:focusy}
Using already obtained grating parameters, in-plane focusing grating coupler is designed that focus the diffracted beam along the direction of propagation (x direction). To focus diffracted beam along transverse direction (y direction), we need to modify the bending of the grating stripes. 

\subsection{Deforming Grating Stripes: Theoretical Formulation}
\label{subsec:bending_stripes}

\begin{figure}[ht!]
		\centering\includegraphics[width=2 in]{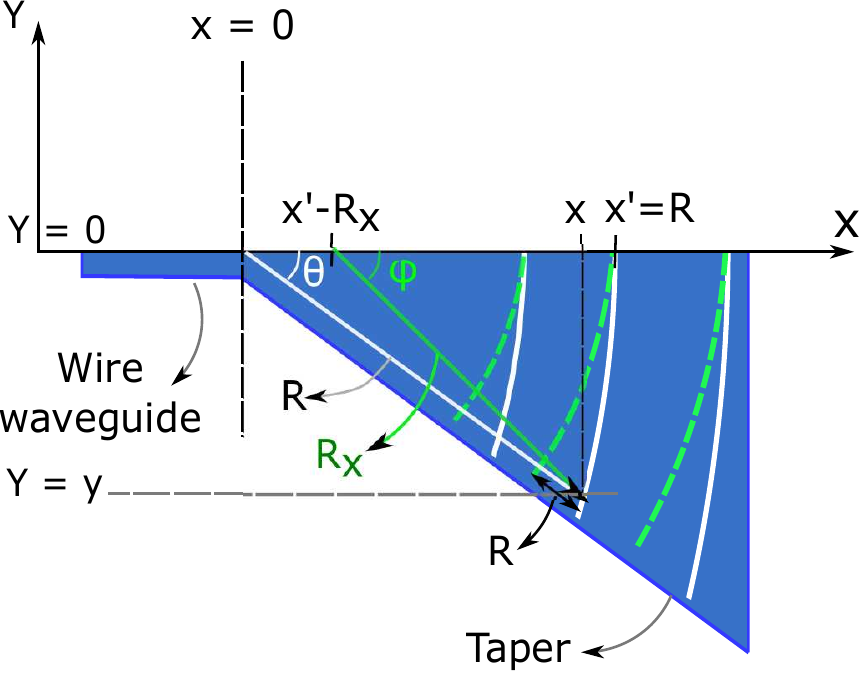}
		\caption{Modulating grating stripes for focusing diffracted beam along y direction. White stripes represents unmodulated and green stripes represent modulated grating periods.}
		\label{fig:yfoc_explanation}
		\end{figure}

Top view of an in-plane focusing grating coupler is shown in fig. (\ref{fig:yfoc_explanation}), for simplicity we look into half section of the coupler which is symmetric with respect to x axis. Propagating mode comes from the wire waveguide and gets expanded in the taper. This beam is diffracted when encounters grating stripes. White lines represent unmodulated grating periods. For focusing along y axis, phase of the diffracted beam at the edges should be leading the phase of beam at the stripe center such a way that it forms a parabola along y axis.    

Simplest way to achieve such functionality is to further bend the grating stripes or reducing the radius of curvature of each period without moving their position as depicted by green stripes compared to their original position presented by white stripes. To obtain a parabolic phase variation along y direction, radius of the stripe has to be reduced by $\Delta R$ such that:

  \begin{equation} \label{eqn:delR}
	 \begin{split}
		\frac{2\pi}{\lambda} n_{eff} \Delta R & =  - \frac{2\pi}{\lambda} \frac{y^2}{2R_y}\\
	  	\Delta R & =  \frac{y^2}{2n_{eff}R_y}  
 	\end{split}
 \end{equation}

In fig. (\ref{fig:yfoc_explanation}), the point (x,y) is located at the edge of a grating stripe. Radius of the unmodulated stripe is R and its origin is located at (0,0). However, when its radius is modified by $\Delta R$ its origin is moved to ($x'-R_x$,0) such a way that ($x'$,0) is the intersection point for both modulated and unmodulated stripe and $R_x$ is the new radius of the stripe. Now the (x,y) point can be described as follows:

  \begin{equation} \label{eqn:def_xy}
	 \begin{split}
		y & = R_x \sin\phi = R sin\theta \\
		x & = R_x \cos\phi + x' - R_x = R_x (\cos\phi -1)+x'	  
 	 \end{split} 	 	 
  \end{equation}
 
where $\phi = \theta + \Delta \theta$ and $\Delta\theta \ll 1$ from which the following can be resolved:

   \begin{equation} \label{eqn:def_delcos}
 	 \begin{split}
 		\Rightarrow \Delta\theta\cos\theta & = (\frac{R}{R_x}-1)\sin\theta	  
  	 \end{split} 	 	 
   \end{equation}

From eqn. (\ref{eqn:def_xy}), x can be elaborated as:

    	 \begin{align*}
			& x = R\cos\theta  = R_x (\cos\theta-\Delta\theta\sin\theta-1) + x'\\ 		
    		\Rightarrow & (R-R_x)(\cos^2\theta - \sin^2\theta)  = (x'-R_x)\cos\theta	  
     	 \end{align*} 	 	 
     
Taking the derivative of the above equation provides a relationship between $dR$ and $dR_x$:

  \begin{equation} \label{eqn:def_dR_dRx}
	 \begin{split}
		\Rightarrow & dR  = (1 - \frac{\cos\theta}{\cos^2\theta-\sin^2\theta}) dR_x
 	 \end{split} 	 	 
  \end{equation}      

Now making an wild approximation that $\theta \Rightarrow 0$, using the property $x = R \cos\theta$ from eqn. (\ref{eqn:delR}) and (\ref{eqn:def_dR_dRx}) we obtain the following:

  \begin{equation} \label{eqn:res_dRx}
	 \begin{split}
		& dR = dR_x (1 - \cos\theta) = - \frac{R^2\sin^2\theta}{2n_{eff}R_y} \\
		\Rightarrow & dR_x = - \frac{R_x^2}{2n_{eff}R_y} (1+\frac{x}{R_x}) 
 	 \end{split} 	 	 
  \end{equation}
  
Eqn. (\ref{eqn:res_dRx}) assumes $R \simeq R_x$ which can be further simplified considering $x \simeq R_x$ resulting in:

	\begin{equation}
		dR_x = - \frac{R^2_x}{n_{eff}R_y}
	\label{eqn:res_dR}
	\end{equation}

If the unmodulated grating periods have a radius $R$, then after focusing modulation to obtain a diffracted phase front of radius $R_y$ the modulated radius ($R'$) of the stripes would be:

 	\begin{equation}
 		R' = R + \Delta R = R (1 - \frac{R}{n_{eff}R_y})
 	\label{eqn:mod_radius}
 	\end{equation}
 	
In simpler words, for focusing light beam in the transverse direction bending of each period has to be modulated according to eqn. (\ref{eqn:mod_radius}) keeping their x-position unperturbed. 	

\section{Simulation Results: Focusing along both direction}
\label{sec:focusboth}
Now this y-focusing transformation is applied to the grating stripes of the in-plane focusing coupler which is described in section \ref{subsec:sim_res_focusx}. For y-focusing design radius was chosen to be $R_y = 150 \mu$m. 

\begin{figure}[ht]
	 	 \centering
         \includegraphics[width=\linewidth]{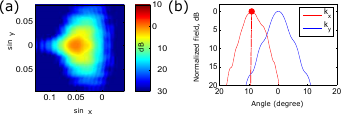}
	 	 \caption{Modeling diffracted beam going upwards for LWD grating coupler at $\lambda = 1550$ nm. (a) Spatial Fourier transform of the near-field measurement. (b) directionality of the diffracted beam, red point marker indicates the angle ($9.23 $ degree) at which maximum is detected.}
	 	 \label{fig:Lens_farfield_focus_b}
	 	 \end{figure}

 Spatial Fourier transform of the near-field measurements ensure the focusing effect of the designed grating coupler in fig. (\ref{fig:Lens_farfield_focus_b}). Diffracted beam radius is 0.107 and 0.099 radians along the x and y direction respectively nearly matching the required fiber mode diameter of $\sim 0.1$ radians (numerical aperture of SMF). Maxima of the focused beam at $\lambda = 1550$ nm would be detected at an angle of $9.23^0$ with the axis perpendicular to the grating surface along x direction (fig. \ref{fig:Lens_farfield_focus_b}(b)). Note that from fig. (\ref{fig:Lens_farfield_apo_b}), we observed beam radius of 0.050 and 0.040 radians along x and y axis respectively in apodized grating which is now matching the fiber mode after implementing focusing modulation along both axial direction.

\begin{figure}[ht]
	 		\centering
            \includegraphics[width=\linewidth]{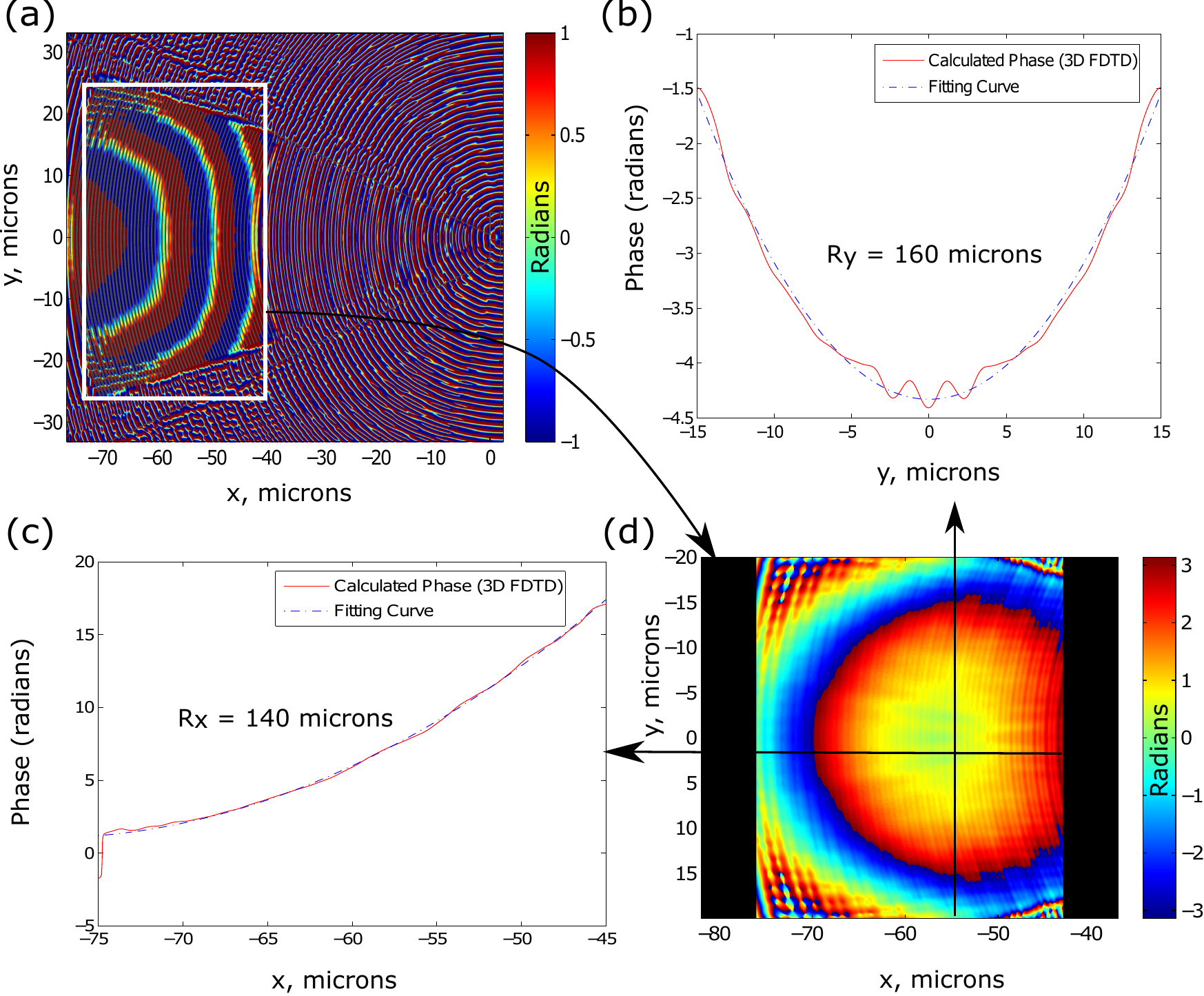}
	 	 		\caption{Phase profile of the LWD grating . (a) Phase profile of the $H_x$ component measured at $z = 0.7 \mu$m above the grating coupler. (b) and (c) show the phase profile along $x = -55 \mu$m and y-axis respectively with second order polynomial fit. (d) Hyperbolic phase front obtained after subtracting linear terms from the phase of $H_x$ inside the white box in (a).}
	 	 	\label{fig:Lens_focusx_3D_phase}
 	 \end{figure}

Spatial Fourier transform of the near-field measurements in fig. \ref{fig:Lens_farfield_focus_b}(a) shows astigmatism effect in the diffracted beam due to mismatch between the radius of curvature of the phase front along two perpendicular axis. However, power in the deformed portion is negligible for the calculated colormap scale is in $dB$. Now it is important to resolve the optimal distance out of plane at which the diffracted beam would match fiber mode size. This distance can be determined from summary of the phase profile depicted in fig. (\ref{fig:Lens_focusx_3D_phase}). Magnetic field component $H_x$ is shown in fig. \ref{fig:Lens_focusx_3D_phase}(a), an white box indicates the portion which encircles the grating region.

We can see the bending of the phase front of $H_x$ above the grating which gives a hint of the phase modulation. Subtracting the linear components from fig. \ref{fig:Lens_focusx_3D_phase}(a), should provide a parabolic phase front above the grating along x and y direction because the diffracted beam has a phase relation as follows:

	\begin{equation}
	   \phi = \frac{2\pi}{\lambda}(\sin\theta) x+ \frac{x^2}{2n_{eff}R^2_x} + \frac{y^2}{2n_{eff}R^2_y}
	   \label{eqn:phase_focusxy}
	\end{equation}

where $R_x$ and $R_y$ are the design radius of curvature along x and y direction and $n_{eff}$ is effective index of the grating. From the color map of phase presented in figure  \ref{fig:Lens_focusx_3D_phase}(d), we can see that the phase front is parabolic shaped in both axial direction. Such phase front shape indicates that the diffracted light beam is focusing.  
	 
 \vspace{0.5em}
	 	 
 To deduce the information where the optimum (focal) point would be located we have extracted the phase front from fig. \ref{fig:Lens_focusx_3D_phase}(d) along the x axis at y = 0 $\mu$m (fig. \ref{fig:Lens_focusx_3D_phase}(c)) and along the y axis at $x = - 55 \mu$m (figure \ref{fig:Lens_focusx_3D_phase}(b)). By fitting the second order polynomial to the extracted phase profile and using eqn. (\ref{eqn:radius_curvature}) radius of curvature along x and y direction are determined to be 137 and 163 $\mu$m respectively. The optimum position for the fiber facet should be between 140 and 160 microns away (out of plane) from the sample surface at an angle of  $9.23^0$ with the z axis. This result is quite consistent with our design parameter $R_x = R_y = 150 \mu$m for phase modulation using eqn. (\ref{eqn:focusx}) and (\ref{eqn:mod_radius}). 
	 	 
\vspace{0.5em}

From the results illustrated above, this could be concluded that a design for grating coupler has been achieved which can focus light out of plane without the aid from external lens. At $\lambda = 1550$ maximum power that can be detected using SMF is $45 \%$ which was calculated using the following overlap integral \cite{taillaert2003efficient} using 2D FDTD simulation:

\begin{equation}
\gamma = \frac{ |\int\int E \times H^*_{fib}|^2}{Re[\int\int E\times H^* \cdot \int\int E_{fib} \times H^*_{fib}]}
\end{equation}

Here $\gamma$ is the overlap between the field and the fiber mode at the focusing point.

\section{Final Remarks}

A simple approach is demonstrated to design long-working-distance (LWD) apodized grating coupler which may find application in efficient fiber coupling in regions that are not easily accessible, in integration and packaging processes. we have derived lensing effect simply modifying the grating periods and bending of the grating stripes. Here we only demonstrate the design method for LWD grating coupler which has a wide scope for development in future.     

\section*{Acknowledgment}
Authors acknowledge financial support of A$\ast$midex Scholarship from Aix-Marseille Universit\'{e} initiative d'excellence and training allowance from Thales Research and Technology, Palaiseau, France.  
\bibliographystyle{IEEEtran}
\bibliography{biblio}

\end{document}